# Designing Ambiguity-Aware Clerical Review: A Stratified Sampling Framework for Record Linkage and Deduplication

Contributions

Joseph Lam, Amaia Imaz Blanco, Efrosini Setakis, Jonny Laidler, Jonny Pearson, Katie Harron, Mario Cortina-Borja, Peter Christen, Giulia Mantovani

1 NHS England, London, United Kingdom

2 UCL Great Ormond Street Institute of Child Health

3 University of Edinburgh/ANU

## Abstract

Clerical review of candidate record pairs remains the de facto gold standard for evaluating record linkage, but is resource-intensive and often designed informally. We propose a design-based framework that treats clerical review as finite-population sampling over fine-grained strata defined by match weight, gamma pattern, record-level ambiguity and demographic group. Bands are constructed from deciles of model-based match probability; within bands, strata are formed by agreement patterns and an ambiguity factor derived from matchability and conditional candidate perplexity. A band-wise margin-of-error profile encodes substantive priorities (e.g. tighter targets in high-score bands), while a single global scaling parameter enforces an overall clerical budget.

From deduplicating a labelled training dataset using Splink (50,000 records; ~478,000 candidate pairs), we compared performance of baseline design reviewing ~23% pairs with "budget" design reviewing ~7% pairs. Under the baseline design, estimates of global and band-specific match rates are accurate and the sampled distributions of comparison patterns, gender and ambiguity track the population reasonably well. Under the budget design, global error roughly doubles, band-level errors increase most in mid-score bands where matches, non-matches and ambiguous cases are intermixed, but accuracy in the highest bands and the band-wise ambiguity profile are largely preserved. Ambiguity-aware allocation thus concentrates clerical effort on records that drive linkage uncertainty, at the cost of reduced representativeness in match patterns and gender.

The framework generalises to other clerical review goals and can incorporate gold standards as priors for design and calibration. It makes explicit, and therefore negotiable, the trade-offs between clerical workload, precision, representativeness and ambiguity coverage.

## 1. Introduction

Clerical matching and reviewing (referred as clerical review in this article) of candidate record pairs remains the de facto proxy gold standard for evaluating record linkage quality (1–3). Human reviewers can identify false matches and false non-matches and provide nuanced judgements for ambiguous cases, often drawing on contextual information that is not available to automated systems, and identify edge cases that connect multiple clusters of records, for example, through address histories, nationality or migration history (4).

However, clerical review is extremely resource-intensive. The time required depends not only on the sheer number of candidate pairs, but also on the information available to reviewers (field content, provenance, metadata (5)), the size and structure of candidate clusters, the design of clerical review software (such as CROW, (6)), and the ambiguity or uniqueness of each record: cases with a clear "winner" among a few candidates are resolved quickly, whereas non-unique, high-ambiguity cases with several plausible candidates are much more costly (3,7).

In large-scale linkage or deduplication, it is typically infeasible to review all candidate pairs. People have used rule-based (8) and sampling-based methods (9,10) to identify record pairs to prioritise for clerical reviewing. Rule-based methods for selecting records for clerical review are difficult to tune to capture informative pairs, and hold strong assumption on the validity of the chosen thresholds. Naïve uniform sampling of potentially non-representative record pairs can misallocate limited expert time and misguide linkage quality evaluation (10). Moreover, not all clerical reviews serve the same purpose: designs for ongoing quality assurance, equity audits, operational "grey-zone" triage, and training of machine-learning linkers tolerate different types and magnitudes of error (Table 1). Strategies to allocate clerical review resource proportional to decision error tolerance is an imminent operational need.

Existing work has formalised pair selection as a probability-sampling problem (11). Dasylva et al. (9) described a framework in which candidate pairs above a score cut-off are stratified by match weight, sampled using Neyman allocation (12) to minimise the variance of global error measures. This approach is markedly more efficient than naïve random sampling but is primarily optimised for a single set of global estimands, with strata defined only by match weights. More recently, the Office for National Statistics' QUAIL project (QUality Analyser for Interpreting Linkage) are exploring Bayesian approaches for candidate stratification and sampling for clerical review (13,14). These developments largely oriented towards global estimands and score-based strata, with limited explicit treatment of equity-focused estimands, operational triage, or ambiguity-aware targeting beyond the match weights.

In this paper, we extend this design-based perspective on clerical review. Our objective is to develop a general framework for selecting record pairs that is (i) grounded in match weights, (ii) sensitive to comparison patterns and demographic strata, and (iii) explicitly informed by record-level ambiguity. Rather than treating clerical review as an ad hoc exercise or a single-purpose precision-recall study, we treat it as a unifying scaffold that can support multiple roles in the linkage pipeline.

| Goal of clerical review | Typical use case | Design priorities | Main estimates | Error tolerance |
|---|---|---|---|---|
| Ongoing deduplication quality assurance | Periodic; when duplications are flagged or reported | Moderate overall budget (e.g. 5–10% of candidate pairs); broadly similar margins across score bands. | Record-based linkage resolution; Temporal changes in band-specific match rates and composition | Very low error tolerance/High relative precision. |
| Equity and subgroup audits | Detecting systematic difference by ethnicity, sex, migration history etc… | Extend stratification to include equity variables (e.g. score band × pattern × subgroup); minimum sample sizes per key subgroup with targeted oversampling of minoritised groups. | Subgroup-specific match rates and between-group differences. | Low error tolerance as between-group comparisons demand a tight error margin in equity strata |
| Clerical triage of ambiguous pairs | Operational review of cases near thresholds; accept automation in clearly low/high bands | Two-tier allocation with dense sampling and tight error margins in "grey-zone"; sparse sampling elsewhere for monitoring only. | Local false match and false non-match rates near operational thresholds, to support operational decision making. | Asymmetric: low error tolerance for false positives at thresholds (e.g. deduplicating spines); high error tolerance in false negatives may be acceptable. |
| Training and re-training of ML linkers | Creating labelled data for supervised or semi-supervised model | Use stratification that prioritises patterns and feature-space coverage | A diverse, rich, labelled training set spanning disagreement patterns, match weights and key demographics. | High error tolerance. The key concern is coverage of important regions of the feature space, assessed via pattern-level metrics. |

*Table 1. Goals of clerical reviews, use case, priorities, estimates and error tolerance*

## 2. Methods

### 2.1 Overview of the framework

The proposed framework has two components:

1. Design of a stratified clerical sampling plan, based on

- (decile) bands of model-based match probability (or match weight);
- within-band match patterns (vectors of agreement/disagreement across identifiers);
- Record-level ambiguity, defined by conditional perplexity and matchability, is used to prioritise how clerical effort is distributed
- an optional demographic stratification (in our case example, gender), to assess robustness and equity of sampled data.

2. Evaluation of the sampling design using data with ground truth.

This is achieved by comparing sample-based estimates of match rates within strata and bands, with the corresponding underlying true rates calculated from the full test population.

Clerical review is framed as a finite-population (hypergeometric) sampling problem. For each design stratum $s$ we specify a design-stage match rate $p_{0,s}$, choose a target margin of error $w_s$for estimating the true stratum match rate $p_s$, derive the required sample size $n_s$, and then assess, how closely the realised estimators recover $p_s$, band-level rates $p_b$, and the global match rate.

The quantities $p_{0,s}$ and $w_s$ are design parameters. They can be chosen to reflect the primary objective of linkage (e.g. conservative deduplication versus population inference) and the acceptable error tolerance across the match probability distribution.

### 2.2 Motivational example: Data and linkage model

Methodological development and evaluation were conducted using the historical_50k dataset distributed with the Splink record linkage library (15) using Python (16) (codes available on GitHub, (17)). The dataset contains 50,000 person records with quasi-identifiers: forename, surname, date of birth (DOB), occupation, birthplace, a synthetic postcode and gender. Each record includes a cluster identifier for the underlying individual, providing the “true” duplicate status of any candidate pair.

We considered a deduplication setting, identifying records belonging to the same individual within a single file. Candidate pairs were generated using blocking rules intended to reflect common operational practice. Blocks combined partial name keys, DOB fragments and postcode prefixes, such as:

- $\text{substring(forename, 1,3)}$and $\text{substring(surname, 1,4)}$;
- exact surname and DOB;

- substring(postcode_fake, 1,3)and DOB.

A probabilistic linkage model was then fitted with the following comparisons:

- A name comparison on a concatenated forename_surname field (forename + surname), with term-frequency adjustment.
- DOB comparison with levels for exact agreement, partial agreement (e.g. year or month only) and disagreement.
- Postcode comparison with levels for exact agreement, partial agreement and disagreement.
- Exact match comparisons on birthplace and occupation, both with term-frequency adjustment.

Field-specific $m$- and $u$-probabilities (agreement probabilities under match and non-match, respectively) were estimated by expectation maximisation (18). For each candidate pair $(i, j)$, the model yields a posterior match probability $\pi_{ij}$. In the present framework, these probabilities are used for ranking and stratification; we do not assume that they are perfectly calibrated to the true match probabilities.

**2.3 Stratification: score bands, comparison patterns and demographic strata**

**2.3.1 Score bands**

To structure clerical review across the score distribution, we partitioned candidate pairs into 10 ordered score bands according to deciles of $\pi_{ij}$. Let $q_k$denote the empirical $(k/10)$-quantile of $\pi_{ij}$. Band $k$is defined as

$$\text{band}_k = \{(i,j): q_{k-1} \leq \pi_{ij} < q_k\}, k = 1, \dots, 10,$$

with band 1 containing the lowest-probability pairs (predominantly true non-matches) and band 10 the highest-probability pairs (predominantly true matches). Intermediate bands correspond to regions of increasing ambiguity. These bands form the primary stratification and allow design-stage precision requirements and clerical effort to vary systematically across the score distribution.

**2.3.2 Comparison-pattern strata**

Within each band, we further stratified by comparison pattern, defined as the vector of discrete comparison levels ("gammas") across fields. Splink encodes comparison outcomes as ordered levels (e.g. 0 = disagreement, 1 = partial agreement, 2 = exact agreement). These levels are concatenated into a pattern code (e.g. 2103x), which uniquely identifies a particular combination of agreements and disagreements across fields (with x denoting missing or unused levels).

For each band, pattern strata are defined as the distinct pattern codes observed among its pairs. In high-score bands, many pairs fall into patterns with agreement on all or most fields; mid-range bands are enriched for "single-field disagreements"; low bands often contain patterns with multiple disagreements.

Each stratum is therefore indexed by:

$$s = (\text{band, comparison pattern}).$$

This second level of stratification ensures that clerical review samples the diversity of disagreement types encountered by the model, rather than only its marginal score distribution.

**2.3.3 Demographic strata**

To probe potential equity issues, we optionally stratified further by demographic variables. In this study we included gender, carried forward into the pair table. For selected bands, particularly high-score bands where automated acceptance may be tempting, we formed strata of the form:

$$s = (\text{band, comparison pattern, gender}),$$

allowing us to assess whether very high-score matches are equally reliable across genders. In principle, the same approach could be applied to other variables (e.g. ethnicity, age or region) depending on the equity questions of interest.

The cross-classification of band, comparison pattern and gender defines the set of design strata. For stratum $s$, we denote:

- $N_s$: number of candidate pairs in the population;
- $M_s = \sum_{(i,j)\in s} Y_{ij}$: number of true matches, where $Y_{ij} = 1$ if $(i,j)$is a true match and 0 otherwise;
- $p_s = M_s/N_s$: true stratum match rate;
- $\pi_s$: mean model match probability within the stratum, used as a design-stage value $p_{0,s}$ for $p_s$.

**2.3.4 Record-level ambiguity: matchability and conditional perplexity**

For each record $i$in the file, the linkage model produces a set of possible counterparts: all records $j$ that form a candidate pair $(i,j)$after blocking. We denote this set by $\mathcal{C}_i$. In addition to these candidates, there is always the possibility that record $i$has no match in the current candidate frame; we treat this as an extra option labelled $j = 0$.

The probabilistic linkage model assigns a weight $w_{ij}$ to each option $j$ in the list "no match + all candidates", that is to each $j \in \{0\} \cup \mathcal{C}_i$. These weights come from the Splink match weight and the chosen prior odds. To turn them into proper probabilities that sum to one, we simply normalise by first compute the total weight

$$W_i = \sum_{k \in \{0\} \cup \mathcal{C}_i} w_{ik},$$

then define the posterior probability of each option as

$$p_{ij} = \frac{w_{ij}}{W_i}, j \in \{0\} \cup \mathcal{C}_i.$$

By construction, these probabilities satisfy $\sum_j p_{ij} = 1$. Intuitively, $p_{ij}$ is the model's belief that record $i$ matches candidate $j$, and $p_{i0}$ is the belief that $i$ has no match in the candidate set. From this, we derive two related measures of record-level ambiguity:

Matchability refers to the posterior probability that record $i$ matches *someone* in the candidate set $m_i = 1 - p_{i0}$; Records with $m_i \approx 1$ are almost certain to have a match; records with $m_i \approx 0$ are almost certainly unmatched (given the candidate set and model).

Conditional perplexity from the probabilities over the candidate matches, conditional on there being a match.

1. **Matchability**

The posterior probability that record $i$ matches *someone* in the candidate set is

$$m_i = 1 - p_{i0}.$$

We refer to $m_i$ as matchability. Records with $m_i \approx 1$ are almost certain to have a match; records with $m_i \approx 0$ are almost certainly unmatched (given the candidate set and model).

2. **Conditional Perplexity**

Conditional on record $i$ having a match, the distribution over candidates is

$$p_{ij}^* = \frac{p_{ij}}{m_i}, j \in \mathcal{C}_i, m_i > 0,$$

with $\sum_{j \in \mathcal{C}_i} p_{ij}^* = 1$. The entropy of this conditional distribution is

$$H_i = -\sum_{j \in \mathcal{C}_i} p_{ij}^* \log p_{ij}^*,$$

and the corresponding **conditional perplexity** is

$$\mathrm{PP}_i = \exp(H_i).$$

Conditional Perplexity $\mathrm{PP}_i$ can be interpreted as the "effective number of plausible candidates" for record $i$, conditional on it having a match. $\mathrm{PP}_i$ values close to 1 indicate a single dominant candidate, while larger values indicate that posterior mass is spread across several candidates.

Therefore, matchability and conditional perplexity $(m_i, \mathrm{PP}_i)$jointly characterise whether a record is likely to match at all, and, if so, how *ambiguous* its candidate set is (perplexity).

**2.3.5 Ambiguity bins and Ambiguity factor**

To obtain an interpretable, ordinal summary of ambiguity for use in design and diagnostics, we grouped records into a small number of ambiguity bins based on $(m_i, \mathrm{PP}_i)$.

We constructed a two-dimensional feature vector. Records with extremely low matchability ($m_i < 0.05$) were treated as effectively non-linkable given the candidate set; they were assigned to the lowest ambiguity level and excluded from the clustering step. On the remaining records, we applied an unsupervised clustering procedure, with the number of clusters $K$constrained to lie between 3 and 6:

$$K \in \{3,4,5,6\}.$$

For each candidate value of $K$, a clustering model was fitted, and an internal model-selection criterion (e.g. information criterion or cluster separation score) was evaluated; the value

$$K^\star = \arg \min_{K \in \{3,\ldots,6\}} \text{criterion}(K)$$

was selected, and the corresponding model used to assign each record to one cluster.

For each cluster $c = 1, \ldots, K^\star$, we computed mean matchability and mean perplexity:

$$\bar{m}_c = \frac{1}{|c|} \sum_{i \in c} m_i, \bar{\mathrm{PP}}_c = \frac{1}{|c|} \sum_{i \in c} \mathrm{PP}_i.$$

Clusters were then ordered by increasing ambiguity, defined as higher perplexity and lower matchability. Specifically, clusters were ranked by the pair, and each record was assigned an ambiguity factor:

$$\text{ambig_factor}_i \in \{1, \ldots, K^\star\},$$

according to the rank of its cluster, with larger values indicating higher ambiguity (more plausible candidates, less concentrated posterior mass). Records with $m_i < 0.05$ were assigned $\text{ambig_factor}_i = 0$. For use in the pair-level analysis, we carried $\text{ambig_factor}_i$ forward to the pair table as an ambiguity bin indicator (ambig_bin) and defined design strata as

$$s = (\text{band,ambiguity bin,comparison pattern,gender}).$$

In both the baseline and budget designs, ambiguity enters the sampling plan in two ways:

1. Stratification – ambiguity bins are part of the stratum definition (band × ambiguity × pattern × gender); and
2. Margin-of-error scaling – within each band, higher-ambiguity bins are assigned relatively tighter target margins of error (and hence larger sample sizes) than lower-ambiguity bins, via an ambiguity-based scaling factor applied to $w_s$.

This makes the design explicitly ambiguity-aware: conditional on score band and comparison pattern, more ambiguous records are preferentially selected for clerical review.

### 2.4 Hypergeometric sampling model and sample size determination

Within each stratum $s$, clerical review draws a simple random sample without replacement of $n_s$ candidate pairs for manual classification. Let $X_s$ denote the number of true matches in the reviewed sample. Conditional on $M_s$, $N_s$ and $n_s$,

$$X_s \mid M_s, N_s, n_s \sim \text{Hypergeometric}\ (N_s, M_s, n_s),$$

and the sample proportion

$$\hat{p}_s = \frac{X_s}{n_s}$$

is an unbiased estimator of $p_s$, with variance

$$\text{Var}\,(\hat{p}_s) = \frac{p_s(1-p_s)(N_s-n_s)}{n_s(N_s-1)}.$$

This equals the binomial variance $p_s(1-p_s)/n_s$ multiplied by the finite population correction $(N_s-n_s)/(N_s-1)$. Using a Normal approximation, a two-sided 95% confidence interval for $p_s$has approximate half-width (margin of error)

$$w_s \approx z_{0.975}\sqrt{\text{Var}\,(\hat{p}_s)} = z\sqrt{\frac{p_s(1-p_s)(N_s-n_s)}{n_s(N_s-1)}},$$

with $z = 1.96$. For design, we substitute a design-stage rate $p_{0,s}$for the unknown $p_s$. Here we set $p_{0,s} = \pi_s$, the mean posterior probability in stratum $s$, truncated away from 0 and 1 to avoid unstable sample-size calculations (e.g. to $[0.05, 0.95]$).

Solving for $n_s$yields the standard hypergeometric sample-size formula:

$$n_s = \frac{z^2 p_{0,s}(1-p_{0,s})N_s}{z^2 p_{0,s}(1-p_{0,s}) + w_s^2(N_s-1)}.$$

We then round $n_s$ up to the nearest integer and truncate at $N_s$, ensuring $0 \le n_s \le N_s$. For very small strata, $n_s = N_s$and the stratum is reviewed exhaustively.

**2.4.1 Baseline margin-of-error profile**

We first specified a baseline band-level margin-of-error profile $w_b^{\text{base}}$, reflecting a deduplication-oriented view that higher-probability bands should be estimated more precisely than low bands. We used:

$$\begin{aligned} w_{b1}^{\text{base}} &= w_{b2}^{\text{base}} = 0.07, \\ w_{b3}^{\text{base}} &= w_{b4}^{\text{base}} = 0.06, \\ w_{b5}^{\text{base}} &= w_{b6}^{\text{base}} = 0.05, \\ w_{b7}^{\text{base}} &= 0.04, \\ w_{b8}^{\text{base}} &= 0.035, \\ w_{b9}^{\text{base}} &= w_{b10}^{\text{base}} = 0.03, \end{aligned}$$

where, for example, 0.03 corresponds to a 3% half-width. For each stratum $s$ in band $b(s)$we set $w_s^{\text{base}} = w_{b(s)}^{\text{base}}$ and calculated $n_s^{\text{base}}$ using the above formula. Within each band, these baseline margins are further modulated by ambiguity: strata in higher ambiguity bins (more ambiguous) receive smaller effective margins and therefore larger sample sizes, while strata in lower ambiguity bins receive larger margins, preserving the overall band-level priority structure while concentrating effort on ambiguous records.

### 2.4.2 Resource-constrained scaling

In practice, total clerical capacity is often capped. To obtain a design with an overall clerical budget fraction $f$ (e.g. 5% of all candidate pairs), we introduced a global scaling factor $c \geq 1$ and defined

$$w_s^{(c)} = c\, w_s^{\text{base}}.$$

For each $c$, we recomputed stratum-specific sample sizes $n_s^{(c)}$ and total sample size $n_{\text{tot}}^{(c)} = \sum_s \; n_s^{(c)}$. For moderate $N_s$, $n_s^{(c)}$ is approximately proportional to $1/\{w_s^{(c)}\}^2$, so $n_{\text{tot}}^{(c)}$ decreases approximately as $1/c^2$.

We used a simple bisection search to choose a global scaling factor $c^\star$. For a given target budget fraction $f$ (here $f = 0.05$), we repeatedly evaluated the total planned sample size $n_{\text{tot}}^{(c)} = \sum_s \; n_s^{(c)}$ and adjusted $c$ until the planned fraction $n_{\text{tot}}^{(c)}/N$ was within 0.1 percentage points of $f$, or until 20 iterations had been reached. This procedure preserves the relative priorities across bands implied by the baseline margin profile while uniformly relaxing all margins just enough to respect the global budget at the design stage.

### 2.5 Estimation from clerical samples

Given a realised clerical sample, we estimate:

- **Stratum-level match rates**

$$\hat{p}_s = \frac{X_s}{n_s},$$

where $X_s$ is the number of true matches in the clerically reviewed sample from stratum $s$.

- **Band-level match rates**
  For band $b$, pooling all strata $s \in b$,

$$\hat{p}_b = \frac{\sum_{s \in b} \; N_s \hat{p}_s}{\sum_{s \in b} \; N_s},$$

i.e. a population-weighted average of stratum estimates.

- **Global match rate**

$$\hat{p}_{\text{global}} = \frac{\sum_s \; N_s \hat{p}_s}{\sum_s \; N_s}.$$

Under the hypergeometric sampling model and assuming error-free clerical classification, these estimators are design-unbiased for $p_s$, $p_b$ and $p_{\text{global}}$, respectively.

### 2.6 Evaluation with synthetic ground truth

#### 2.6.1 True rates and sample-based model

Because cluster identifiers are known, the **true** match rate in stratum $s$ is

$$p_s^{\text{true}} = \frac{1}{N_s} \sum_{(i,j)\in s} Y_{ij},$$

and the true band-level and global rates are

$$p_b^{\text{true}} = \frac{1}{N_b} \sum_{(i,j)\in b} Y_{ij}, p_{\text{global}}^{\text{true}} = \frac{1}{N} \sum_{(i,j)} Y_{ij},$$

where $N_b$ is the number of pairs in band $b$ and $N$ is the total number of candidate pairs.

In each simulation replicate, we draw a clerical sample according to the stratified design and treat the known $Y_{ij}$ as the outcome of (hypothetical) clerical review. The sample-based stratum model is

$$\hat{p}_s = \frac{1}{n_s} \sum_{(i,j)\in s,\ \text{sampled}} Y_{ij},$$

which defines a piecewise-constant function that can be applied to all pairs by their stratum membership.

#### 2.6.2 Performance metrics

We assess the sampling performance in two ways.

1. **Global performance**

$$E_{\text{global}} = | \hat{p}_{\text{global}} - p_{\text{global}}^{\text{true}} |$$

2. **Band-level performance**
   For each band $b$,

$$E_b = | \hat{p}_b - p_b^{\text{true}} |,$$

are computed for each replicate and reported in percentage points.

### 2.6.3 Representativeness metrics

To quantify how well the clerical sample reflects heterogeneity within bands, we examine the L1 distance between the population and sample distributions of comparison patterns and gender. For band $b$, let $p_b^{\text{pop}}(k)$be the population proportion of category $k$(pattern or gender) and $p_b^{\text{samp}}(k)$the corresponding sample proportion. The L1 distance is

$$\mathrm{L1}_b = \sum_k \mid p_b^{\text{pop}}(k) - p_b^{\text{samp}}(k) \mid \in [0,2],$$

and we report

$$\mathrm{L1}_b^{\text{pct}} = 50 \times \mathrm{L1}_b \in [0,100],$$

interpretable as the percentage of probability mass that would need to be reallocated for the sample to match the population distribution. We compute $\mathrm{L1}_{b,\text{pattern}}^{\text{pct}}$and $\mathrm{L1}_{b,\text{gender}}^{\text{pct}}$ separately for each band.

### 2.6.4 Ambiguity coverage metrics

For each record $i$, we compute matchability $m_i$, conditional perplexity $\mathrm{PP}_i$, and the ambiguity factor as defined in Section 2.3.4. We then derive the following summaries:

1. **Band-level mean ambiguity**
   For each band $b$, we compute population and sample means:

$$\bar{m}_b^{\text{pop}}, \bar{m}_b^{\text{samp}}, \bar{\mathrm{PP}}_b^{\text{pop}}, \bar{\mathrm{PP}}_b^{\text{samp}},$$

and report absolute differences $\mid \bar{m}_b^{\text{samp}} - \bar{m}_b^{\text{pop}} \mid$and $\mid \bar{\mathrm{PP}}_b^{\text{samp}} - \bar{\mathrm{PP}}_b^{\text{pop}} \mid$.

2. **Band-wise composition by ambiguity factor**

We tabulate the composition of each band by ambiguity level, both in the population and in the clerical sample. We then compare $p_{b,a}^{\text{samp}}$and $p_{b,a}^{\text{pop}}$across ambiguity levels within each band. These summaries allow us, for each band, to see whether the clerical sample is concentrated on low-ambiguity (“easy”) records or whether it retains an appropriate share of higher-ambiguity records that are more likely to drive linkage uncertainty.

### 2.6.5 Replication and assumptions

The sampling and estimation procedure is repeated over multiple random seeds, yielding empirical distributions for $E_{\text{global}}$ and $E_b$ and allowing both typical and worst-case performance to be characterised.

Interpretation rests on the following assumptions:

1. Clerical decisions as gold standard
   In simulation, $Y_{ij}$ is treated as error-free. Real clerical decisions are subject to misclassification; this is not modelled here.

2. Homogeneity within strata
   Match rates within each stratum is assumed sufficiently homogeneous that a single rate $p_s$ is meaningful for design and estimation.

3. Approximate design rates
   Sample sizes are computed using $p_{0,s} = \pi_s$. In general, $p_{0,s} \neq p_s^{\text{true}}$. Robustness to misspecification of $p_{0,s}$ is assessed empirically in the simulations.

## 3. Results

### 3.1 Pair-level Score bands and match probabilities

The scored candidate set included $N = 478{,}307$ pairs. By construction, pairs were partitioned into ten decile bands of model-based match probability (b1–b10), with band sizes ranging from 34,736 (b2) to 55,466 pairs (b1) (Figure 1, Table 2). As expected, the empirical true match rate increased across bands: the lowest bands contained almost exclusively non-matches (0.17% in b1; 0.96% in b2), while the highest band (b10) consisted almost entirely of true matches (99.99%). Intermediate bands showed a gradual transition from mostly non-matches (4.5% in b3; 12.0% in b4) to predominantly matches in the upper bands (72.5% in b7, 89.2% in b8).

| band_10 | count | mean match probability (sd) | true match rate (%) |
|---|---|---|---|
| b1 | 55466 | 0 (0) | 0.17 |
| b2 | 34736 | 0 (0) | 0.96 |
| b3 | 45752 | 0 (0) | 4.51 |
| b4 | 51605 | 0 (0) | 12.00 |
| b5 | 47801 | 0 (0) | 26.96 |
| b6 | 53107 | 0.02 (0.02) | 71.35 |
| b7 | 52566 | 0.15 (0.1) | 72.54 |
| b8 | 46281 | 0.71 (0.16) | 89.21 |
| b9 | 45819 | 0.98 (0.02) | 98.16 |
| b10 | 45182 | 1 (0) | 99.99 |

*Table 2. Score-band descriptives for candidate pairs: count, mean match probability and true match rate. Sd = standard deviation*

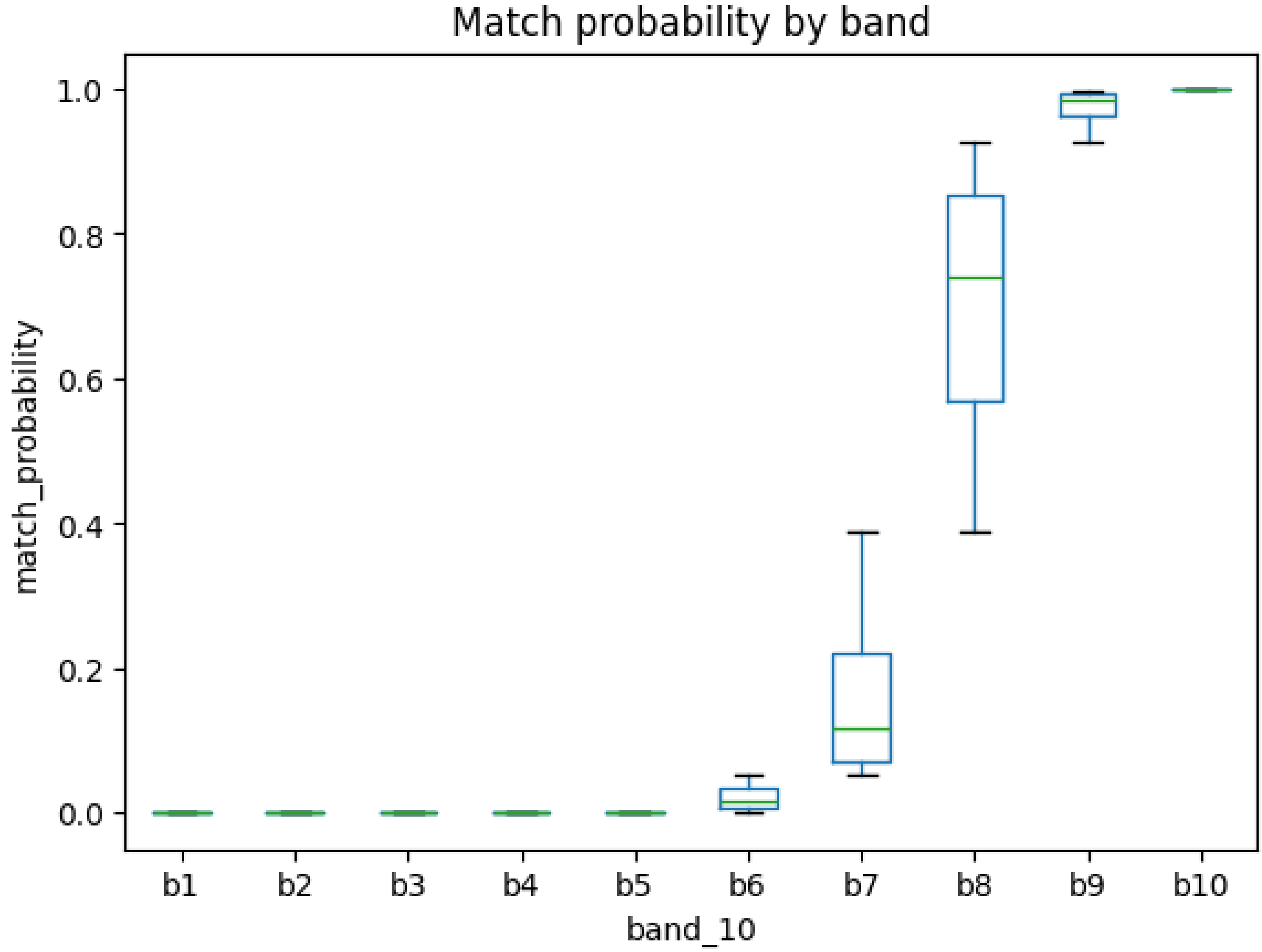

*Figure 1. Box chart of match probability by band decile.*

### 3.2 Record-level ambiguity: matchability and conditional perplexity

Matchability and conditional candidate perplexity (CCP) are defined at the record level. To characterise record-level ambiguity, we first compute for each record $i$its matchability $m_i$and conditional perplexity $\mathrm{PP}_i$, and then cluster records on $(m_i, \mathrm{PP}_i)$into six ambiguity bins (a1–a6), as described in Section 2.3.5 (Table 3). Bins a1 and a2 correspond to structurally “easy” records: a1 contains records with very low matchability and near-unit perplexity (essentially no plausible match), whereas a2 contains records with high matchability and low perplexity (a single dominant candidate). Bins a3 and a4 have high matchability but increasing perplexity (on average 2.4 and 5.1 effective candidates, respectively), representing records that almost certainly match but with multiple plausible candidates. Bin a5 combines low matchability with moderate perplexity, and a6 captures a small set of extremely ambiguous records with very high perplexity and modest matchability, corresponding to dense local clusters where candidate identity is highly uncertain. See examples of records in each ambiguity bin in Supplementary Table 1.

| bin | type | *n* | mean CCP (sd) | mean matchability (sd) |
|---|---|---|---|---|
| a1 | low matchability, low perplexity | 29,952 | 1.07 (0.24) | 0.06 (0.10) |
| a2 | high matchability, low perplexity | 16,716 | 1.13 (0.19) | 0.95 (0.10) |
| a3 | high matchability, moderate perplexity | 10,985 | 2.44 (0.50) | 0.96 (0.09) |
| a4 | high matchability, moderate perplexity | 5,936 | 5.13 (1.40) | 0.94 (0.10) |
| a5 | moderate matchability, moderate-high perplexity | 3,016 | 5.30 (1.99) | 0.15 (0.16) |
| a6 | low matchability, very high perplexity | 293 | 76.59 (104.07) | 0.20 (0.27) |

*Table 3. Ambiguity bins based on matchability and conditional candidate perplexity (CCP), defined at the record level. Sd = standard deviation.*

### 3.3 Matchability–perplexity plots for high-score bands

Figure 2 display the joint distribution of matchability and CCP for records whose candidate pairs fall in bands b6–b10. We restrict the plot to these bands because conditional perplexity is only informative once there is a non-trivial probability of matching: in the lowest bands (b1–b5) most pairs correspond to structurally obvious non-matches with very low match probability. Within bands 6–10, most records with high-scoring candidates (especially those contributing to b9–b10) cluster at matchability close to 1 and CCP close to 1, consistent with clear single-candidate matches. At the same time, there is a visible tail of records with elevated perplexity in the upper-mid bands (b6–b8), corresponding to ambiguity bins a3–a6. These ambiguous records are precisely those that the ambiguity-aware sampling design aims to prioritise for clerical review. Supplementary Figure 1 describes the matchability-perplexity plot for all 10 bands.

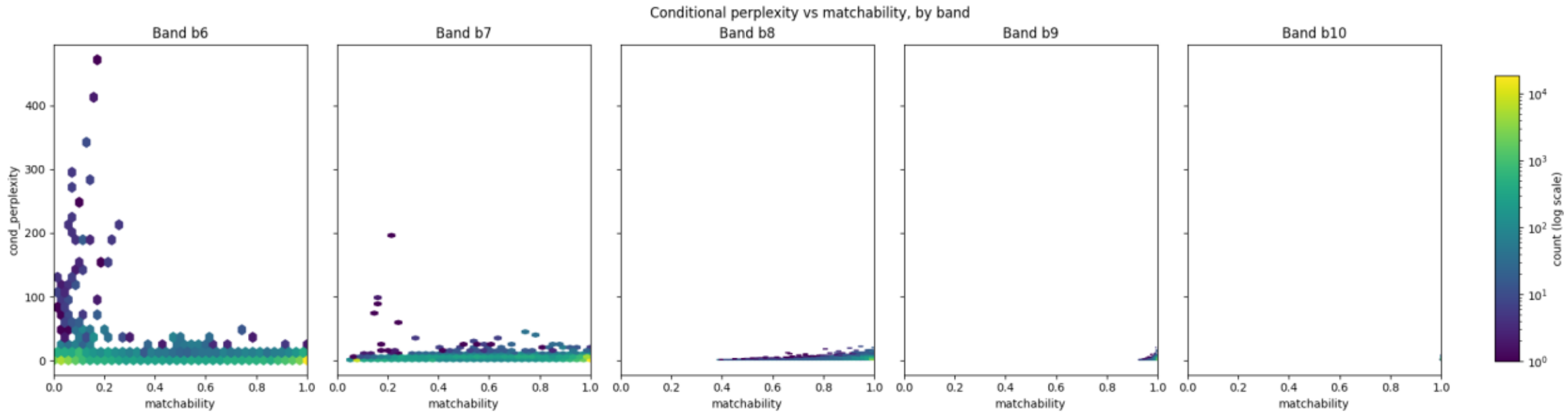

*Figure 2. Conditional perplexity by matchability, by band for bands 6-10. colour indicates the log count of examples (dark blue = few, yellow = many). Band b6 contains many low-matchability, high-perplexity examples, while higher bands progressively shift mass toward low perplexity and higher matchability, with bands b9–b10 dominated by well-matched, low-perplexity points.*

### 3.4 Comparison of baseline and budget designs

#### 3.4.1 Global and band-specific error

To satisfy the specific margin of error, the baseline design sampled on average 109,479 pairs (22.9% of all candidate pairs) per replicate, whereas the budget design sampled 34,263 pairs (7.2%), corresponding to an approximate two-thirds reduction in clerical workload. Although the budget design was calibrated to a nominal 5% target at the design stage, the realised sample fraction in this dataset was 7.2%, reflecting discretisation and approximation in the sample-size calculations.

Across five simulation replicates, the mean absolute deviation between the estimated and true global match rate was 4.24% under the baseline design. Under the budget design, the mean absolute global error increased to 9.32% (Supplementary Table 2 for global error rates, and Supplementary Table 3 for band-level deviation from truth).

At the band level, the increase in error was heterogeneous and concentrated in the mid-range bands where ambiguity is greatest (Figure 3, left). Under the baseline design, mean absolute errors ranged from essentially zero in the highest band (b10: 0.01%) to around 11.4% in b5. Errors were small in the extreme bands (1.1% in b1; 0.71% and 0.45% in b8 and b9), moderate in b2 and b6–b7 (~5%), and largest in the mixed bands b3–b5 (7.3%, 8.2% and 11.4%, respectively).

Under the budget design, band-level errors increased most sharply in these mixed bands (b3-b5). Mean absolute error rose to 13.5%, 18.4% and 23.3% in b3, b4 and b5. Errors in b6–b7 approximately doubled, while the low and high bands remained well controlled: b1 increased to 3.5%, and b8–b10 remained below 1.2%.

These patterns are consistent with the intended priority structure. The ambiguity-aware budget design preserves high accuracy in the upper bands (b8–b10), which are the most relevant for conservative acceptance or deduplication, while accepting substantially larger errors in the mid-score bands (b3–b7) where the mixture of matches, non-matches and ambiguous cases makes estimation inherently difficult.

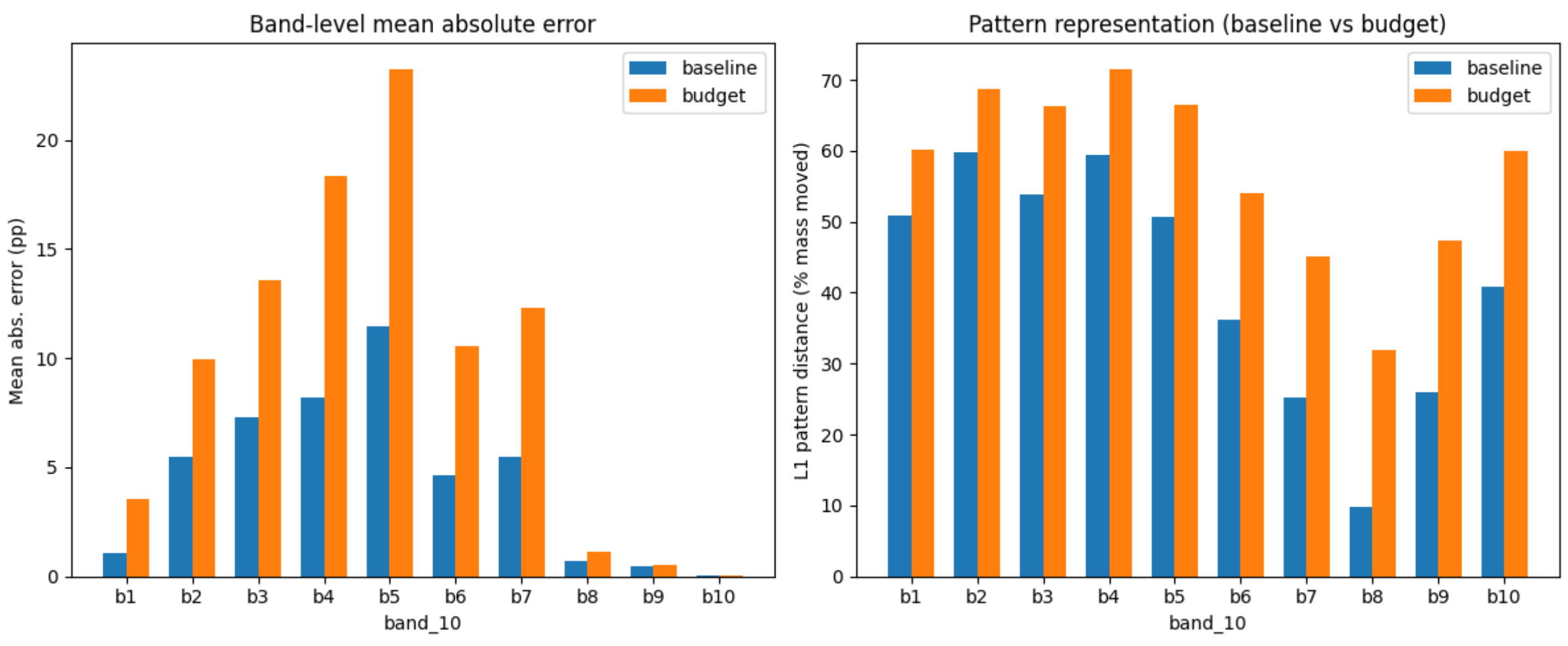

*Figure 3. Band-level mean absolute error by band (left), and pattern L1 distance by band (right) for both baseline and budget design.*

### 3.4.2 Representativeness by comparison pattern, gender and ambiguity

We next compared how well the baseline and budget designs preserved the internal composition of each band in terms of comparison patterns, gender and ambiguity. Representativeness was quantified using the L1 distance between the population and sample distributions, expressed as the percentage of mass that would need to be reallocated for the two distributions to coincide.

Under the baseline design, pattern L1 distances varied widely across bands, from around 10% in b8 to approximately 50–60% in several bands (e.g. b1, b2, b4) (Figure 3, right). This reflects the fact that, even with a relatively generous sampling fraction, the full diversity of comparison patterns is only partially captured in bands with many rare patterns.

Under the baseline design, gender L1 distances were consistently smaller, typically between 3% and 13%, indicating that the sampled gender distribution remained close to the population distribution (Figure 4). Ambiguity L1 distances were generally modest (around 1–26%, Figure 5), with larger values primarily in bands with low sampling fractions and more heterogeneous ambiguity structure.

Under the budget design, representativeness deteriorated for comparison patterns and gender, but changed much less for ambiguity. Pattern L1 distances increased in every band, for example, from 40.8% to 59.9% in b10, from 26.0% to 47.3% in b9, and from 36.2% to 54.0% in b6. Gender L1 distances also increased systematically by 4%–11%, reflecting a moderate loss of fidelity in the band-wise gender mix when the sample was reduced.

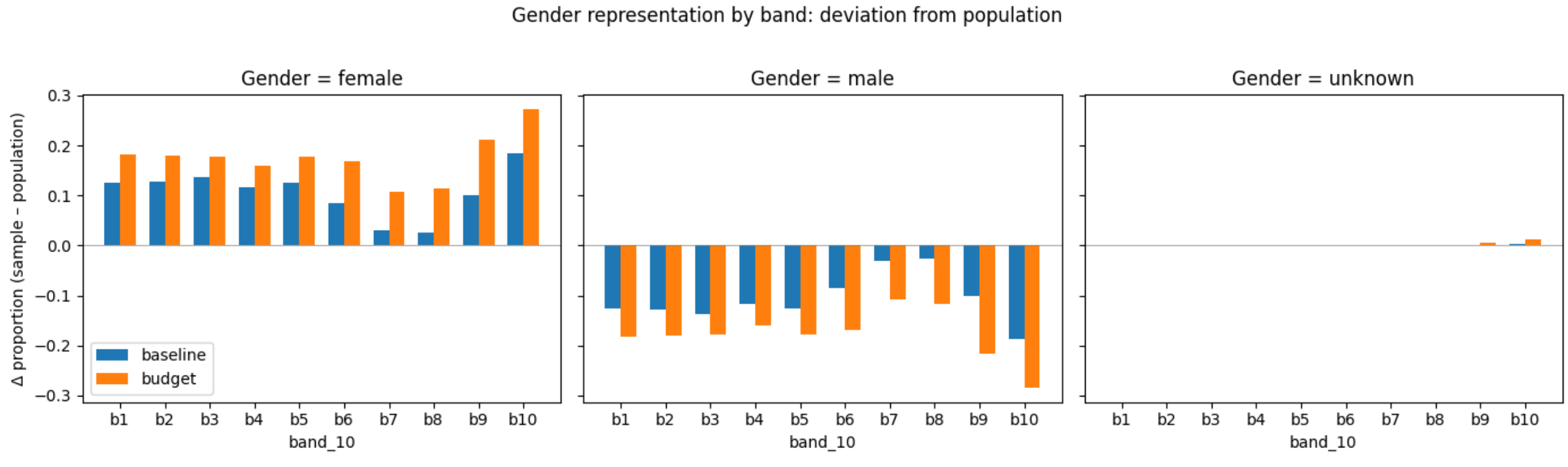

*Figure 4. Deviation of gender representation from population, by gender, for baseline and budget design*

By contrast, changes in ambiguity L1 distances were small (Figure 5). for each ambiguity bin (a1–a6), the proportion of pairs in that bin across score bands in the population and in the baseline and budget samples. For the main matchable bins (a2–a4), both designs closely track the population distribution, particularly in the higher bands where these records are concentrated. For the highly ambiguous bins (a5–a6), both designs retain non-trivial coverage in the mid-score bands (b3–b5) where these cases occur, and the budget design does not systematically under-sample them relative to the baseline. This supports the interpretation that the ambiguity-aware allocation preserves coverage of difficult cases even under a reduced global budget.

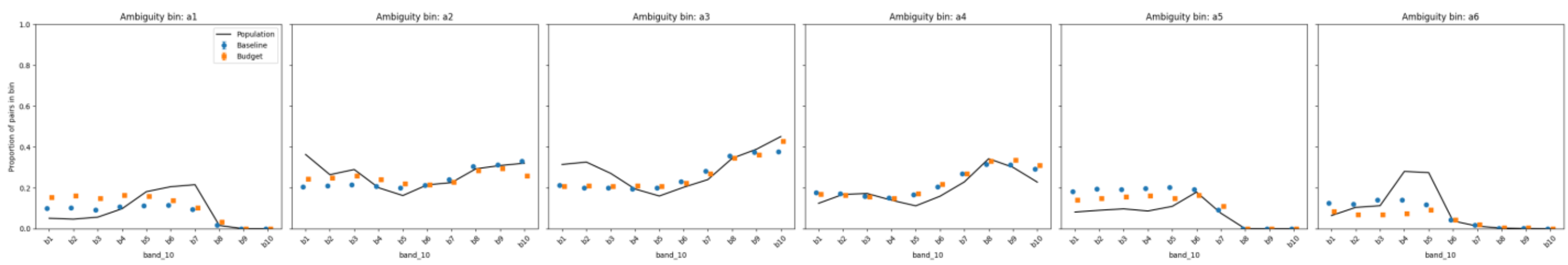

*Figure 5. Comparing representativeness of sample ambiguity across bins, in baseline and budget design.*

## 4. Discussion

The proposed framework combines (i) a hypergeometric sampling model over fine-grained strata defined by score band, comparison pattern, record-level ambiguity and demographic group; (ii) a band-wise margin-of-error profile that encodes substantive priorities; and (iii) a global scaling parameter that enforces an overall clerical budget.

In the example deduplication setting, a baseline design reviewing ~23% of candidate pairs delivered accurate estimates of global and band-specific match rates and reasonably representative samples of comparison patterns and gender, particularly in the higher score bands. A budget design (~7% of pairs) preserved good accuracy in the highest bands (b8–b10) and maintained the ambiguity profile of the clerical sample, but at the cost of larger errors and reduced representativeness in the mid-score bands where matches, non-matches and ambiguous cases are most intermixed. These patterns mirror the practical tensions faced in large-scale linkage systems.

Unlike Neyman allocation (9), which minimises the variance of a single global error metric for a fixed sample size, our design fixes band- and stratum-specific margins of error and uses a global scaling factor to reconcile these with a given clerical budget, prioritising local precision (e.g. in high-score or high-ambiguity strata) over optimality for a single aggregate estimand. An equity-aware linkage quality evaluation depend on how the model behaves locally: around thresholds, within specific demographic groups, and for small sets of ambiguous, high-impact records, rather than on a single global error rate. By encoding band-wise and subgroup-specific error tolerances directly into the sampling design, the proposed framework complements existing global-error approaches and supports more targeted allocation of clerical effort to the regions where mistakes are most consequential.

In many linkage settings, high-quality truth is available for restricted portions of the data: cohorts with strong identifiers (e.g. NHS numbers, social security numbers) or workflows that historically relied on full manual verification. Within the proposed framework, such trusted sub-cohorts are useful in two ways. First, band- and pattern-specific match rates estimated in these sub-cohorts can supply informative design-stage values $p_{0,s}$ when planning clerical sampling in less certain regions. Rather than relying solely on model-based probabilities or diffuse defaults, priors for $p_{0,s}$ can be anchored in empirical experience from high-quality data, improving sample-size calculations in sparse or noisy strata. Second, the same priors create an explicit target for calibration. When the framework is applied to less well-identified cohorts, deviations between observed clerical outcomes and priors derived from trusted sub-cohorts, particularly when stratified by demographic group or  ambiguity, highlight where the linkage model behaves differently in noisier data. Systematic departures (for example, consistently lower match rates in particular bands) can signal model misspecification, instability of score distributions or structural differences in data quality.

**Limitations and future work**

Several limitations should be acknowledged. Clerical decisions are treated as error-free, whereas in practice they are subject to misclassification and may vary across reviewers or over time, and clerical matching errors likely to occur disproportionately for records with higher ambiguity (9). The assumption of approximate homogeneity within each (band, match pattern, demographic) stratum is pragmatic rather than exact, although it is made explicit and can be probed using this framework. A further practical limitation is that the framework requires tuning multiple variance and calibration parameters by band, rather than a single global variance.

Despite these limitations, the proposed framework provides a transparent, extensible way to design and evaluate clerical review strategies under realistic resource constraints. By making explicit the trade-offs between clerical workload, estimation error, representativeness and ambiguity coverage, it supports more principled decision-making about where scarce human effort is best spent in large-scale linkage and deduplication projects.

**Practical guidance for pipeline designers**

The framework proposes a design language that pipelines can adopt for different clerical tasks, making explicit what error structures are being accepted in exchange for feasible levels of human review. A practical recipe for real systems is:

1. **Specify the clerical review goals and decisions**
2. **Translate decision tolerances into target errors** at the relevant level (global, band, subgroup).
3. **Choose a baseline margin profile** $w_b^{\text{base}}$ that reflects these priorities (e.g. very small $w_b$ in bands feeding auto-accept rules, larger elsewhere).

4. **Use the hypergeometric formulas** to compute stratum sample sizes and a baseline total workload.
5. **If needed, scale to a feasible budget** via the global factor $c^{\star}$, and check whether the induced errors remain within tolerable bounds.
6. **Revisit the design iteratively**, tightening margins in bands or strata where errors are too large, loosening them where precision is more than sufficient.

**Conclusion**

We have proposed a design-based framework for clerical review that treats the problem as finite-population sampling over rich strata defined by score band, comparison pattern, record-level ambiguity and demographic group. By specifying band-wise margins of error and enforcing a global budget via a single scaling parameter, the framework makes explicit the trade-offs between clerical workload, accuracy and representativeness. In doing so, it supports more transparent and principled decisions about where scarce human effort is best allocated in large-scale record linkage and deduplication.

Supplementary Materials

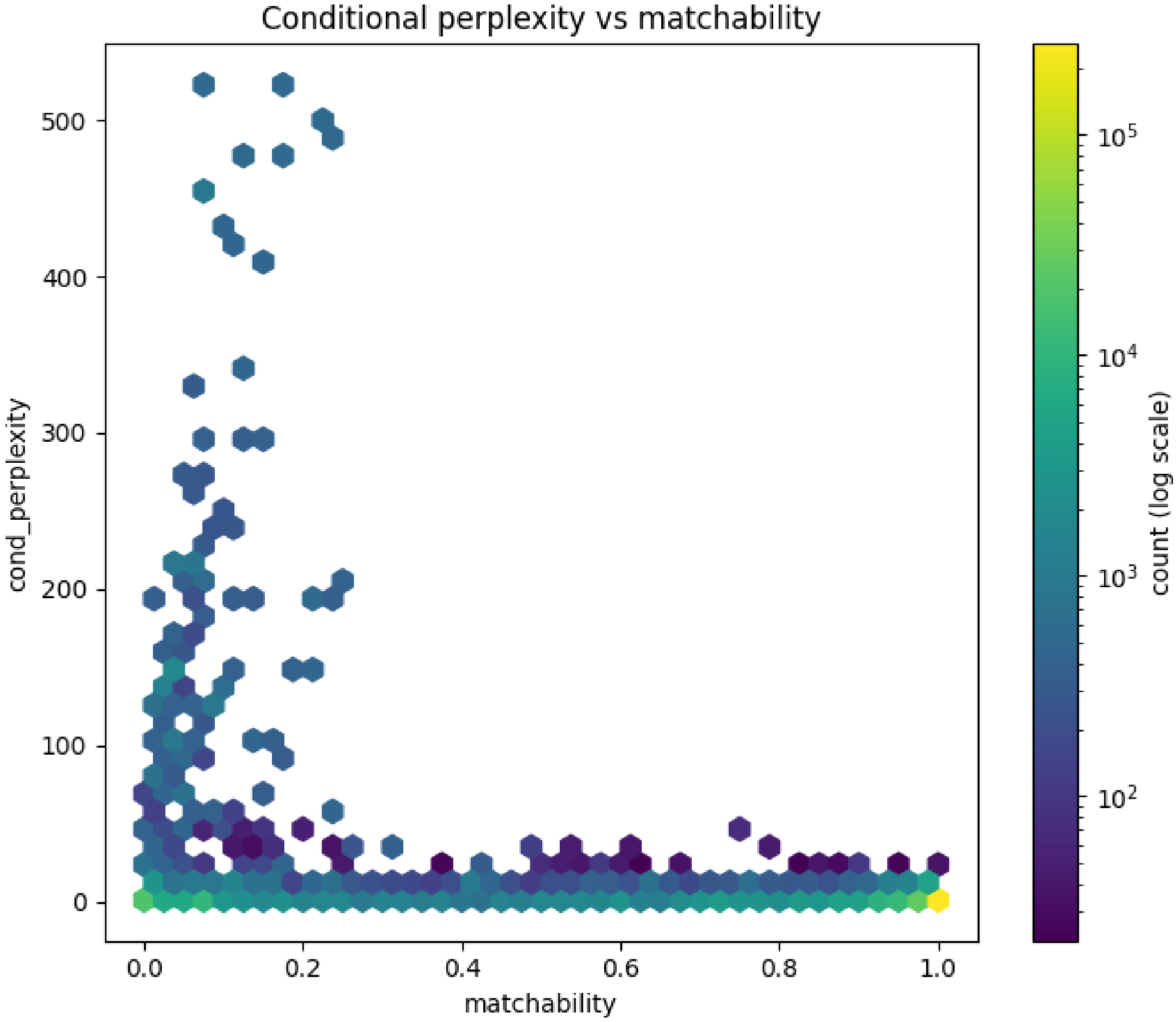

Supplementary Figure 1: Conditional perplexity by matchability for the entire population.

Supplementary Table 1

Supplementary Table 2.

| | mean_global_abs_error (%) | max_global_abs_error (%) | mean_n_sample |
|---|---|---|---|
| baseline | 4.24 | 4.28 | 109479 |
| budget | 9.32 | 9.36 | 34263 |

Mean Global deviation from ground truth across 5 iterations (%)

Supplementary Table 3

| band_10 | true_match_rate (%) | mean_error_base (%) | mean_error_budget (%) | delta_mean_error (%) |
|---|---|---|---|---|
| b1 | 0.17 | 1.06 | 3.53 | 2.47 |
| b2 | 0.96 | 5.47 | 9.97 | 4.5 |
| b3 | 4.51 | 7.31 | 13.54 | 6.23 |
| b4 | 12 | 8.16 | 18.36 | 10.2 |
| b5 | 26.96 | 11.44 | 23.26 | 11.82 |
| b6 | 71.34 | 4.6 | 10.53 | 5.93 |
| b7 | 72.54 | 5.45 | 12.33 | 6.88 |
| b8 | 89.21 | 0.71 | 1.11 | 0.4 |
| b9 | 98.16 | 0.45 | 0.54 | 0.09 |
| b10 | 99.99 | 0.01 | 0.01 | 0.01 |

Band-level deviation from truth, budget versus baseline, across 5 iterations and percentage differences